\documentclass[aps,reprint,longbibliography,superscriptaddress]{revtex4-2}
\usepackage{mathrsfs}
\usepackage{amsmath,gensymb}
\usepackage{amsfonts}
\usepackage{amssymb}
\usepackage{amsthm}
\usepackage{graphicx}
\usepackage{natbib}
\usepackage{color}
\usepackage{hyperref}
\usepackage{bm}
\usepackage[caption=false]{subfig}
\usepackage{verbatim}
\usepackage[normalem]{ulem}
\usepackage{soul}

\begin{document}
\title{Gate-controlled proximity effect in superconductor/ferromagnet van der Waals heterostructures}

\author{G. A. Bobkov}
\affiliation{Moscow Institute of Physics and Technology, Dolgoprudny, 141700 Moscow region, Russia}

\author{K. A. Bokai}
\affiliation{St. Petersburg State University, 7/9 Universitetskaya nab., 199034 St. Petersburg, Russia}
\affiliation{Moscow Institute of Physics and Technology, Dolgoprudny, 141700 Moscow region, Russia}

\author{M. M. Otrokov}
\affiliation{Instituto de Nanociencia y Materiales de Arag\'on (INMA), CSIC-Universidad de Zaragoza, Zaragoza 50009, Spain}

\author{A. M. Bobkov}
\affiliation{Moscow Institute of Physics and Technology, Dolgoprudny, 141700 Moscow region, Russia}

\author{I.V. Bobkova}
\affiliation{Moscow Institute of Physics and Technology, Dolgoprudny, 141700 Moscow region, Russia}
\affiliation{National Research University Higher School of Economics, 101000 Moscow, Russia}

\begin{abstract}
The discovery of 2D materials opens up unprecedented opportunities to design new materials with specified properties. In many cases, the design guiding principle is based on one or another proximity effect, i.e. the nanoscale-penetration of electronic correlations from one material to another. In a few layer van der Waals (vdW) heterostructures the proximity regions occupy the entire system. Here we demonstrate that the physics of magnetic and superconducting proximity effects in 2D superconductor/ferromagnet vdW heterostructures is determined by the effects of interface hybridization of the electronic spectra of both materials. The degree of hybridization can be adjusted by gating, which makes it possible to achieve a high degree of controllability of the proximity effect. In particular, we show that this allows for electrical switching of superconductivity in such structures on and off, as well as for control of the amplitude and sign of the Zeeman splitting of superconducting spectra, opening interesting opportunities for spintronics and spin caloritronics.
\end{abstract}

\maketitle

\section{Introduction}
It it  well-known that in superconductor/ferromagnet (S/F) heterostructures the exchange field of the F layer suppresses the superconductivity in the S layer near the S/F interface via the proximity effect\cite{Buzdin2005,Bergeret2005}. The mechanism of the suppression is twofold. First, the F induces triplet correlations in the S at the expense of singlet ones thus suppressing the singlet superconducting order parameter. Second, if the F is metallic the Cooper pairs can penetrate into it and be destroyed there by the ferromagnetic exchange field. In a thin-film S/F bilayer, where the thicknesses of the S and F layers, $d_S$ and $d_F$, are small with respect to the superconducting and ferromagnetic coherence lengths $\xi_S$ and $\xi_F$, respectively, the suppressing action of these two factors can be described by modeling the bilayer by a homogeneous superconducting film with a reduced effective coupling constant $\lambda_{eff} = \nu_S d_S \lambda/(\nu_S d_S + \nu_F d_F)$ in an effective reduced exchange field $h_{eff} = \nu_F d_F h/(\nu_S d_S + \nu_F d_F)$. Here $\lambda$ is the superconducting coupling constant of the isolated S layer and $h$ is the exchange field of the isolated ferromagnet, $\nu_{S(F)}$ is the density of states at the Fermi level in the S (F) layer \cite{Bergeret2001}. With increasing of the  effective exchange field the superconducting order parameter $\Delta$ is monotonically suppressed \cite{Sarma1963}. The behavior of the electron density of states (DOS) in the effective exchange field $h_{eff}$ is also well-studied and has a typical BCS-like shape with Zeeman splitting of the coherence peaks \cite{Bergeret2018review}. The splitting of the peaks is $2h_{eff}$. 

All physics discussed above is related to bilayers, which are thin with respect to the coherence lengths, but still contain large number of atomic layers, such that the properties of the S and F layers are close to the properties of the corresponding bulk materials and effects of an interface hybridization of electronic spectra \cite{Cardoso2023,Russmann2023}  or quantum-size effects \cite{Shanenko2006,Shanenko2007,Chen2012} are not important. However, the advent of 2D materials has provided unprecedented opportunities for novel heterostructures in the form of van der Waals (vdW) stacks, laterally stitched 2D layers and more complex architectures\cite{Geim2013,Novoselov2016,Castellanos-Gomez2022}. 
Nowadays a lot of vdW materials, which retain their superconducting \cite{Ge2015,Lu2015,Xi2016,delaBarrera2018,Lu2018} and ferromagnetic \cite{Gibertini2019} properties up to the monolayer thickness are being discovered, thus enabling exploration of the proximity effects in the S/F heterostructures in the truly 2D limit.

\begin{figure}[tb]
	\begin{center}
		\includegraphics[width=85mm]{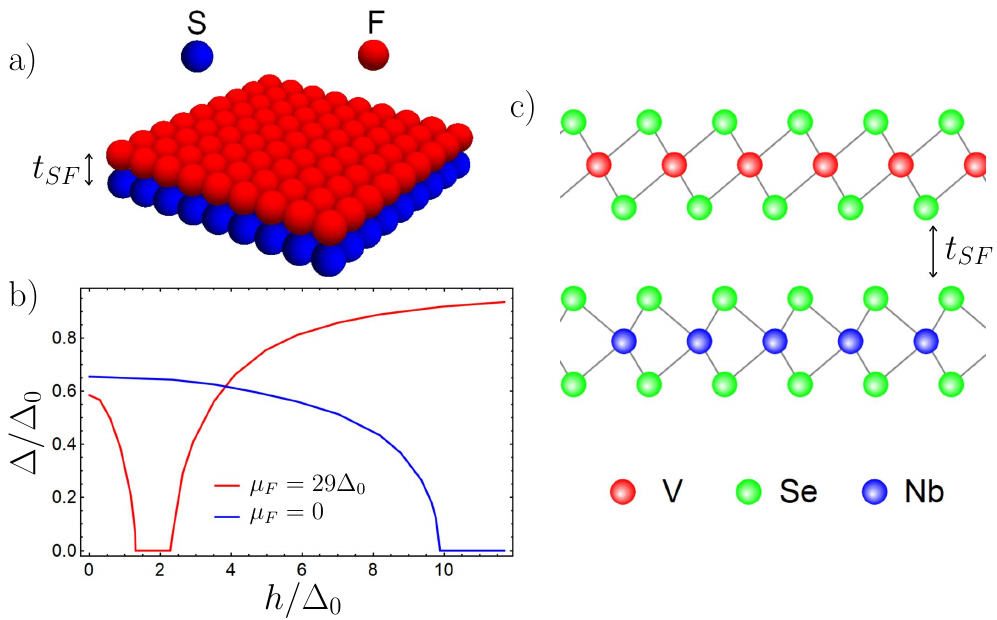}
\caption{(a) Model S/F bilayer with square atomic lattice. $t_{SF}$ is the hopping element between the S and F layers. (b) $\Delta(h)$ for two different values of $\mu_F$ calculated for the model bilayer from panel (a). The red curve corresponds to the case of strong hybridization of electronic spectra of S and F layers, and the blue curve illustrates the case of weak hybridization. $t_S = 70 \Delta_0$, $t_F = 1.25t_S$, $\mu_S=23\Delta_0$. $t_{SF}=1.2\Delta_0$ and $t_{SF}=6\Delta_0$ for red and blue curves, respectively. $T=0.35\Delta_0$. $\Delta_0$ is the superconducting order parameter of the isolated S layer at the same temperature. (c) Atomic structure of the $\mathrm{NbSe_2/VSe_2}$ bilayer.
}
 \label{fig:sketch}
	\end{center}
 \end{figure}

\begin{figure}[tb]
	\begin{center}
		\includegraphics[width=80mm]{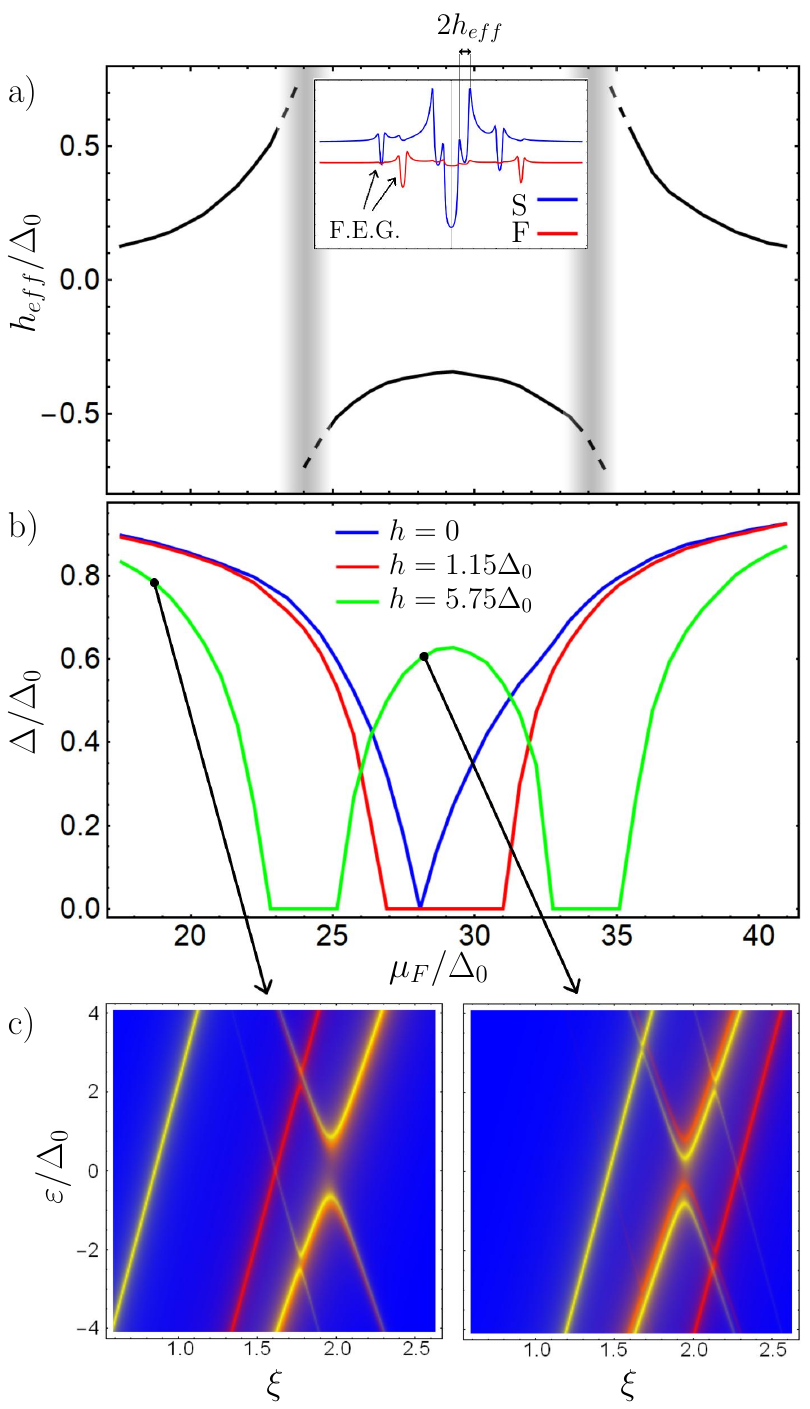}
\caption{All the results are obtained in the model of the square lattice bilayer. (a) $h_{eff}(\mu_F)$ extracted from the Zeeman splitting of the DOS. In the grey regions we are not able to extract  reliable data. Inset: DOS as a function of quasiparticle energy. Blue and red curves represent the DOS in the S and the F layers, respectively. The Zeeman splitting of the DOS $2h_{eff}$ is shown. F.E.G. are finite energy gaps appearing as a result of the S and F layers spectra hybridization. (b) $\Delta(\mu_F)$ for different values of the exchange field $h$ of the F layer. (c) Spectral function of the S/F bilayer. Quasiparticle spectra $\varepsilon(\xi)$ for spin-up (yellow) and spin-down (red) quasiparticles are seen. Left picture corresponds to $\mu_F = 18.8\Delta_0$ and for the right picture $\mu_F = 28.3\Delta_0$. Other parameters are the same as in Fig.~\ref{fig:sketch}.}
 \label{fig:gating_mu}
	\end{center}
 \end{figure}

Proximity effects are particularly relevant to 2D materials since they can alter the electronic properties of an entire material. Works on different proximity effects in vdW S/F heterostructures are already actively appearing, and interest in such heterostructures is growing. In particular, in a F/$\mathrm{NbSe_2}$/F heterostructure, the interaction between the Ising superconductivity in $\mathrm{NbSe_2}$ and magnetism was predicted to control the magnetic ordering of the heterostructure \cite{Aikebaier2022}, and a giant anisotropy of magnetoresistance was observed in Ising superconductor/ferromagnetic insulator junctions \cite{Kang2021}. The interplay between spin-orbit coupling, ferromagnetism and superconductivity in  $\mathrm{NbSe_2/CrBr_3}$ heterostructures was studied \cite{Wickramaratne2021}. Infinite magnetoresistance and nonreciprocal charge transport in $\mathrm{NbSe_2/CrSBr}$ superconducting spin
valves were investigated \cite{Jo2023}, reemergent superconductivity in $\mathrm{NbSe_2/CrCl_3}$ structures was found \cite{Jiang2020}. Topological superconductivity was observed in vdW heterostructures combining a two-dimensional ferromagnet with a $\mathrm{NbSe_2}$ superconductor \cite{Kezilebieke2020}. Also, Josephson effect in S/F/S vdW heterostructures was investigated \cite{Ai2021,Idzuchi2021}.  Here we  study the magnetic proximity effect in a S/F vdW heterostructure made of atomically thin S and F materials, as exemplified by 1H-$\mathrm{NbSe_2}$ and 1T-$\mathrm{VSe_2}$, respectively. We find that the magnetic proximity effect in the superconductor including the induced Zeeman field and the resulting suppression of  superconductivity is determined by the hybridization of the electronic bands of S and F and can be dependent on the strength of the ferromagnet in a very nontrivial way.

2D materials possess unique features, which allow for high degree of controllability of superconductivity \cite{Ueno2008,Ye2010,Zeng2018,Saito2016,Lu2018,Zhang2013} and ferromagnetism \cite{Deng2018,Matsuoka2023,Tan2021} by gating. The magnetic proximity effect in nonsuperconducting vdW heterostructures \cite{Zhong2017,Norden2019,Tang2020,Ghiasi2021,Wu2021} was also reported to be tuned by gating \cite{Cardoso2023}. Here we predict that  2D S/F heterostructures also allow for high controllability of the proximity effects. It is demonstrated that the proximity influence of the F layer on the superconductivity is strongly sensitive to the difference between the filling factors of the S and F layers. This sensitivity is due to the fact that in such 2D heterostructures the particular value of $h_{eff}$, which is induced in the S layer, and the particular degree of a Cooper pair destruction due to the leakage in the F are determined by the degree of hybridization of electronic spectra of the individual materials. 
Remarkably, the degree of this hybridization can be varied by changing the relative filling factors via gating, which enables tuning of the Zeeman splitting, including reversal of its sign, opening interesting opportunities for spintronics and spin caloritronics. The superconducting order parameter can also be tuned. Moreover, the superconductivity can be  purposefully switched off and on. It is worth noting that superconductivity here does not simply play the role of an indicator of induced Zeeman splitting. Recent experimental and theoretical studies have demonstrated a rich variety of transport phenomena occurring in devices based on Zeeman-split superconductors \cite{Bergeret2018review}. In particular, Zeeman split superconductivity allows heat-to-spin conversion to be performed with the highest possible efficiency. This fact leads to the emergence of giant thermoelectric and thermospin effects \cite{Machon2013,Machon2014,Ozaeta2014}, which open up the possibility of highly efficient thermally induced motion of domain walls in S/F heterostructures \cite{Bobkova2021}. Furthermore, the simultaneous presence of the Zeeman splitting and strong spin-orbit coupling allows for non-unitary pairing and possibility for dissipationless spin transport \cite{Bobkov2024_spin}. Thus, our results show that vdW S/F heterostructures can become a promising platform for the creation of electrically controlled 2D Zeeman-split superconductors with great perspectives in thermoelectricity and low-dissipative spintronics. We begin by considering the basic physics in the framework of a simplified tight-binding model and then demonstrate the same physics in the heterostructure composed of real vdW materials.

\section{Basic physics: bilayer in the framework of a tight-binding hamiltonian on a square lattice} 
System under consideration is shown in Fig.~\ref{fig:sketch}(a). It consists of two 2D layers: a 2D F and a 2D S. The system is modeled by the following tight-binding Hamiltonian on a square lattice:
\begin{align}
&\hat H = \sum\limits_{i,\alpha,\beta}\hat c^\dagger_{i,\alpha}\left(\begin{matrix}0&0\\0&(\bm h\bm\sigma)_{\alpha,\beta}\\\end{matrix}\right)\hat c_{i,\beta}-
\sum\limits_{i,\sigma}\hat c^\dagger_{i,\sigma}\left(\begin{matrix}\mu_S&0\\0&\mu_F\\\end{matrix}\right)\hat c_{i,\sigma} \nonumber \\
&+\sum\limits_i\left[\hat c_{i,\uparrow}\left(\begin{matrix}\Delta&0\\0&0\\\end{matrix}\right)\hat c_{i,\downarrow}+H.c.\right]- \sum\limits_{<ij>,\sigma}\hat c^\dagger_{i,\sigma}\left(\begin{matrix}t_S&0\\0&t_F\\\end{matrix}\right)\hat c_{j,\sigma} \nonumber \\
&-\sum\limits_{i,\sigma}\hat c^\dagger_{i,\sigma}\left(\begin{matrix}0&t_{SF}\\t_{SF}&0\\\end{matrix}\right)\hat c_{i,\sigma}
\label{eq:hamiltonian}
\end{align}
Here $\hat c_{i,\sigma} = (c_{i,\sigma}^S, c_{i,\sigma}^F)^T$ is a vector composed of annihilation operators for electrons belonging to the S and F layers at site $i$ in the plane of each layer and for spin $\sigma = \uparrow, \downarrow$. $\mu_{S,F}$ are on-site energies of the S and F layers, respectively, which determine the filling factors of the conduction band of the materials and in case of isolated layers correspond to their chemical potentials. We assume only nearest-neighbor hopping and $t_{S,F}$ is the nearest-neighbor hopping element in the planes of the S and F layers. $\langle ij \rangle$ means summation over nearest neighbors. $t_{SF}$ is the hopping element between the S and F layers. $\bm h$ is the exchange field of the ferromagnet. $\Delta$ is the superconducting order parameter in the superconductor, which is to be calculated self-consistently as $\Delta = \lambda \langle c_{i,\downarrow}^S c_{i,\uparrow}^S \rangle$, where $\lambda$ is the pairing constant. The superconducting order parameter and the DOS are calculated via the Green's functions methods, see Appendix \ref{app:A} for details of the method.

Fig.~\ref{fig:sketch}(b) represents the dependence of the superconducting order parameter on the exchange field $h = |\bm h|$ of the F layer for two different values of $\mu_F$ and the same $\mu_S$. 
Blue curve in Fig.~\ref{fig:sketch}(b) corresponds to the case when $\mu_S$ and $\mu_F$ differ strongly. This dependence $\Delta(h)$ is just a suppression of the superconductivity by the proximity to the F layer, which is very similar to the classical result obtained for thin-film 3D S/F bilayers for not very low temperature $T$ \cite{Sarma1963}. Turning to the red curve in Fig.~\ref{fig:sketch}(b), which corresponds to closer values of $\mu_F$ and $\mu_S$, we see that $\Delta(h)$ is qualitatively different. At first $\Delta$ is completely suppressed by $h$ and then is restored practically to the value of the isolated S layer. That is, in the considered case of the 2D S/F bilayer the proximity effect can be nontrivial and the particular behavior depends strongly on the difference between the chemical potentials of the layers. Below we demonstrate that the physics of the proximity effect is determined by the hybridization of the electronic spectra of the S and F layers. For the blue curve the hybridization of the electronic spectra of S and F layers is weak, and for the red curve it is strong. The degree of the hybridization is controlled by the difference between the chemical potentials.

In principle, we can apply to the 2D bilayer our intuition developed by studying the proximity effect in conventional 3D thin-film S/F bilayers. As it was mentioned in the introduction, the proximity influence of the F layer on superconductivity is described by an effective exchange field $h_{eff}$ induced in the S layer and by the destruction of the Cooper pairs in the F layer.  $h_{eff}$ can be extracted from the Zeeman splitting of the DOS. In the considered 2D S/F bilayer the DOS also has typical BCS-shape with the Zeeman splitting. An example is shown in the inset to Fig.~\ref{fig:gating_mu}(a). The effective exchange field determined from the Zeeman splitting of the DOS is plotted in Fig.~\ref{fig:gating_mu}(a) as a function of $\mu_F$. It is seen that it very strongly and non-trivially depends on the chemical potential. {\it Not only the amplitude but also the sign of $h_{eff}$ can be adjusted by gating}, what can be of interest for spintronics applications. In the grey regions in Fig.~\ref{fig:gating_mu}(a) it is difficult to correctly extract $h_{eff}$ from the DOS, see below. The dependence of the superconducting order parameter $\Delta$ on $\mu_F$ is demonstrated in Fig.~\ref{fig:gating_mu}(b). We can see that $\Delta(\mu_F)$ is fully correlated with $h_{eff}(\mu_F)$: the larger the absolute value of $h_{eff}$ the stronger the superconductivity is suppressed. However, it is worth noting that not only $h_{eff}$ determines the suppression of $\Delta$. The ``leakage'' of Cooper pairs into the F layer and their subsequent destruction there also contributes. We get back to this point below upon discussing the  $\mathrm{NbSe_2/VSe_2}$ bilayer structure.

Now we discuss physical mechanism of the high and non-trivial sensitivity of the proximity effect, presented in Figs.~\ref{fig:gating_mu}(a)-(b), to the chemical potential. 
The electronic spectra $\varepsilon(\xi)$ of the S/F bilayer corresponding to the parameters of the left black point at the green curve of Fig.~\ref{fig:gating_mu}(b) are shown in the left picture of Fig.~\ref{fig:gating_mu}(c).  In the considered model the electronic spectra depend on the momentum only via $\xi = -2(\cos p_x a + \cos p_y a)$, where $a$ is a lattice constant. For this reason in Fig.~\ref{fig:gating_mu}(c) the energy is shown as a function of $\xi$ instead of more convenient for first-principle calculations representation as a function of the momentum along certain directions in the Brillouin zone (BZ). Two spin-split electronic branches originating from the F layer and two parabolic BCS-like electronic branches coming from the S layer are clearly seen. The S-branches are also spin split and this is a signature of the presence of $h_{eff}$ in the superconductor. The F and S branches are hybridized in the form of anti-crossing at the points of intersection of the branches belonging to the same spin. These anti-crossings result in the finite-energy gaps (F.E.G.), which are also seen in the DOS, see inset to Fig.~\ref{fig:gating_mu}(a). For a chosen value of $\mu_F$ the F-branches and S-branches are far enough from each other and, therefore, their hybridization is rather weak resulting in the rather small value of the spin splitting of the S-branches and $h_{eff}$. Upon increasing the value of $\mu_F$ we are moving to the right along the green curve in Fig.~\ref{fig:gating_mu}(b). In this case, the F-branches of the spectrum will move closer to the S-branches and the degree of their hybridization grows. This results in the growth of the spin-splitting of the S-branches and $h_{eff}$ and the stronger  suppression of superconductivity. When one of the F-branches comes as close as possible to the S-branches, the hybridization is maximal, the spectra are strongly reconstructed and the superconductivity is fully suppressed. In this region of $\mu_F$ the correct determination of $h_{eff}$ is not possible due to the strong reconstruction of the spectra. 

Upon further increase of $\mu_F$  the spin-down F-branch already ``passed through the S-branches'' and moved some distance away from them, as it is seen from inspection of the right picture of Fig.~\ref{fig:gating_mu}(c). As a result the hybridization becomes weaker, the absolute value of $h_{eff}$ decreases and the superconductivity is restored. It is interesting that in this region of $\mu_F$ the effective exchange field $h_{eff}$ changes sign, as it is seen from Fig.~\ref{fig:gating_mu}(a) and also from electronic spectra in the right picture of Fig.~\ref{fig:gating_mu}(c), where the spin-down (red) branches are above the spin-up (yellow) branches unlike the left picture of Fig.~\ref{fig:gating_mu}(c). The reason is that, in fact, the F- and S-branches cannot ``pass through each other'' due to the nature of the anti-crossing process. 
Namely, at certain minimal separation in $\xi$, they abruptly switch around due to a strong hybridization, i.e., the leftmost (rightmost) spin-down branch with a positive slope, which originally belonged to F (S), now becomes the S-branch (F-branch). 
As a result, all the evolution of the electronic spectra of the S/F bilayer upon varying of $\mu_F$ obeys the rule: the order of alternating spin-up and spin-down branches must be preserved. For this reason the sign of $h_{eff}$ is reversed. It is reversed again upon further increase of $\mu_F$ when the second F-branch goes across the S-branches.

 \begin{figure}[tb]
	\begin{center}
		\includegraphics[width=85mm]{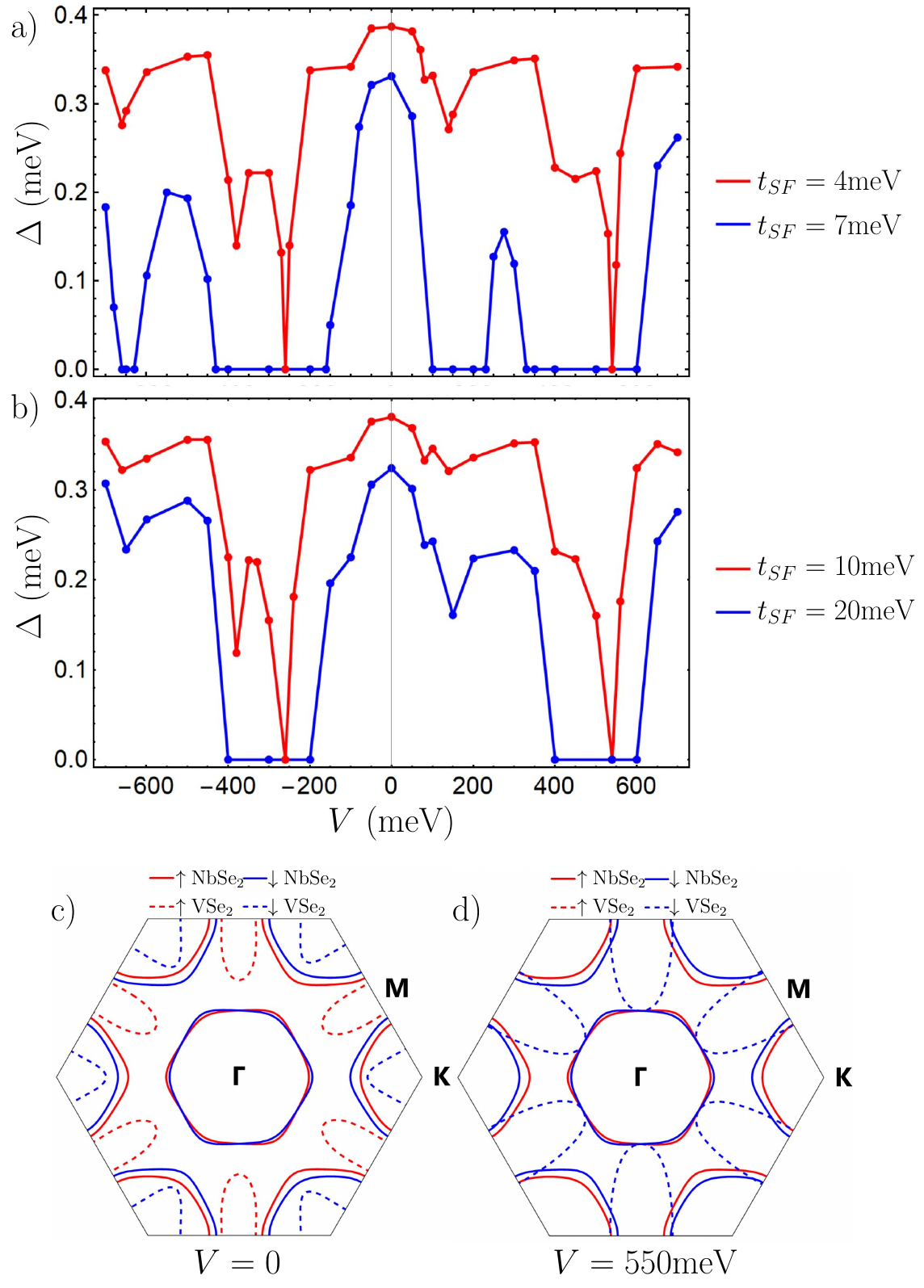}
\caption{$\mathrm{NbSe_2/VSe_2}$ bilayer. (a)-(b) Superconducting order parameter as a function of gating potential $V$ of the F layer for (a) OOP and (b) IP orientations of $\mathrm{VSe_2}$ magnetization. $T=2$K, critical temperature of the monolayer $\mathrm{NbSe_2}$ was chosen $T_c=2.7$K, which is similar to the experimental value. (c)-(d) Spin-split Fermi surfaces obtained from  single-band tight-binding hamiltonians of the $\mathrm{NbSe_2}$ (solid) and $\mathrm{VSe_2}$ (dashed) monolayers with no gating potential in $\mathrm{VSe_2}$ (c) and at $V=550$meV (d). Please note that electronic structure of $\mathrm{VSe_2}$ contains an additional Fermi-pocket around the $\Gamma$-point, see Appendix~\ref{app:B}. However, it is not fitted by our single-band hamiltonian and is not shown here because it is far from the $\mathrm{NbSe_2}$ Fermi-surfaces and thus is not hybridized with them.}
 \label{fig:real_delta}
	\end{center}
 \end{figure}

 \begin{figure}[tb]
	\begin{center}
		\includegraphics[width=85mm]{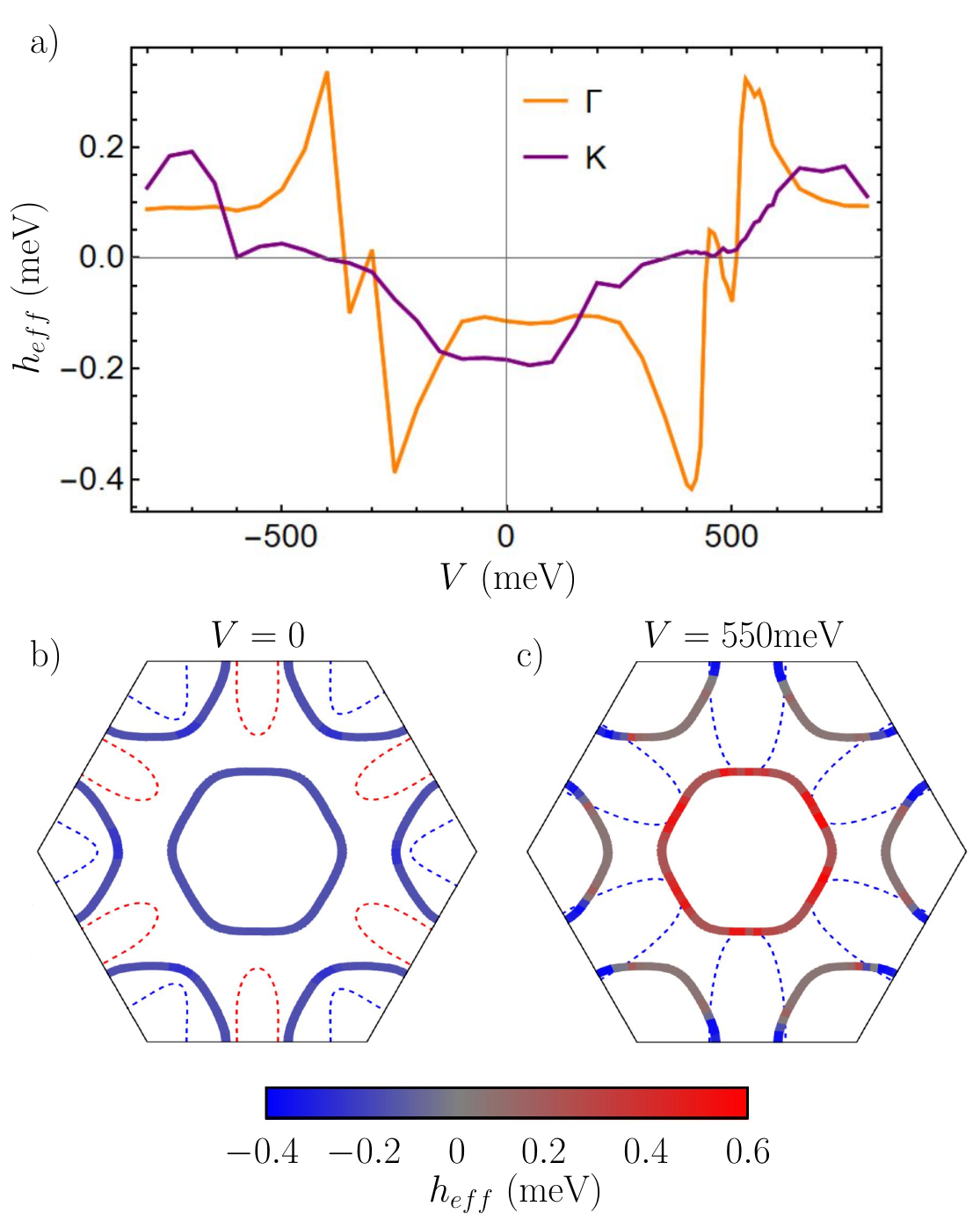}
\caption{$\mathrm{NbSe_2/VSe_2}$ bilayer, OOP magnetization orientation. (a)  $h_{eff}^{\Gamma }$ extracted from the splitting of the coherence peaks of the momentum-averaged DOS at the Fermi surface around $\Gamma$-point (orange curve) and $h_{eff}^{\mathrm K}$ extracted from the sum of the DOS around all $\mathrm K$-points (purple). (b-c) The effective exchange field $h_{eff}^{\Gamma (\mathrm K)}(\bm n)$ for a given momentum direction $\bm n = \bm p/|\bm p|$ and for a given Fermi surface around the $\Gamma (\mathrm K)$-point at $V=0$ (b) and $V=550$meV (c). All the parameters of the bilayer are the same as in Fig.~\ref{fig:real_delta}. }
 \label{fig:real_zeeman}
	\end{center}
 \end{figure}

\section{$\mathrm{NbSe_2/VSe_2}$ bilayer structure} 
Therefore, the high tunability of the proximity effect in the 2D S/F bilayers is determined by the hybridization of the electronic spectra of the materials composing the bilayer. A fair question arises: to what extent does this effect depend on the specific model describing the bilayer and whether it will be preserved for real vdW materials, the electronic spectra of which, as a rule, are not at all described by the considered model? To answer this question we chose a monolayer 1H-$\mathrm{NbSe_2}$ as superconductor \cite{Xi2016,delaBarrera2018,Khestanova2018,Wickramaratne2023} and a monolayer 1T-$\mathrm{VSe_2}$ as a ferromagnet \cite{Ma2012,Bonilla2018,Yu2019,Chua2020,Huang2023} to investigate the proximity effect in the S/F bilayer constructed from these materials. 
The atomic structure of the
heterostructure is illustrated in Fig.\ref{fig:sketch}(c). 
We examine the electronic structure of both the isolated $\mathrm{NbSe_2}$ and $\mathrm{VSe_2}$ monolayers and the $\mathrm{NbSe_2/VSe_2}$ bilayer using the density functional theory (DFT) electronic structure calculations and use their results to build an approximate normal state single-band tight-binding hamiltonian of the $\mathrm{NbSe_2/VSe_2}$ heterostructure, see Appendix \ref{app:B} for details.

The value of the interlayer hopping $t_{SF}$, extracted by fitting the DFT-calculated electronic spectrum of the  $\mathrm{NbSe_2/VSe_2}$ heterostructure, was estimated to be $t_{SF} = 30$meV. 
It results in extremely strong proximity effect, which fully suppresses superconductivity. We believe that this value is an upper theoretical limit of the interlayer hopping for ideally matching lattices of the materials in a heterobilayer. In real heterostructures it should probably be several times smaller. Moreover, if materials with lattice constants that are not close to each other are chosen for the experimental implementation of a heterostructure, this  leads to a further decrease in the effective interlayer hopping parameter. For these reasons we demonstrate in Fig.~\ref{fig:real_delta} the results for smaller $t_{SF}$. 

Panels (a) and (b) of Fig.~\ref{fig:real_delta} represent superconducting order parameter for the $\mathrm{NbSe_2/VSe_2}$ heterostructure as a function of the shift in $\mu_F$ produced by gating, which  is denoted as $V$. It is worth noting that $V$ does not coincide with the true gate potential applied experimentally, which can be several times larger\cite{Jiang2018,Ye2010}. In real experimental situation $\mu_S$ is also shifted due to the gating, but its shift is much smaller than $V$ due to the Debye screening of the potential. Actually, for the considered effect only the difference $\mu_F - \mu_S$ is important. For this reason we neglect the small shift of $\mu_S$. First of all, it worth noting that unlike our toy model, for the $\mathrm{NbSe_2/VSe_2}$ bilayer we observe a strong anisotropy of the proximity effect depending on the orientation of the $\mathrm{VSe_2}$ magnetization $\bm M$. Panel (a) corresponds to the out-of-plane (OOP) orientation of $\bm M$ and panel (b) shows results for the in-plane (IP) orientation. The overall suppression of superconductivity is much stronger for the OOP orientation. This fact is well-known and originates from the Ising-type anisotropy of the $\mathrm{NbSe_2}$ \cite{Xi2016,Saito2016_SUST,Dvir2018,Sohn2018,delaBarrera2018}.  Because of this for the OOP orientation we considered smaller values of $t_{SF}$ than for the IP orientation. The same $t_{SF}$ values as for the IP orientation result in almost full suppression of superconductivity in case of OOP orientation. 

Experimental and numerical data on the magnetic anisotropy in $\mathrm{VSe_2}$ monolayer are controversial, as both types of anisotropy were reported in the literature \cite{Bonilla2018,Chua2020}. Nevertheless, we found that the controllability of the proximity effect is achievable in both  cases, the basic physics being the same as in our toy model. Gating of the F layer shifts its electronic spectrum near the Fermi surface and thus varies degree of  hybridization between NbSe$_2$ and VSe$_2$.  Since the full picture of electronic spectra is rather complicated in this case, in Figs.~\ref{fig:real_delta}(c)-(d) we demonstrate the Fermi surfaces of the heterostructure with no gating $V=0$ and at $V=550$meV, respectively. The plotted Fermi surfaces correspond to $t_{SF}=0$ (separate S and F layers), but on the scale of these figures the interlayer hopping gives only very weak modifications of the curves.  In Fig.~\ref{fig:real_delta}(c) it is seen that at $V=0$ the Fermi surfaces originating from the F ans S layers are far from each other, the hybridization of the spectra is weak and the suppression of superconductivity is also minimal. In Fig.~\ref{fig:real_delta}(d) the Fermi surfaces intersect, the hybridization has the strongest value and the suppression of superconductivity is maximal. 

The weaker suppression of superconductivity for IP orientation is connected to the fact that the spectra of the S layer are protected from exchange spin splitting by the Ising spin-orbit effective field and $h_{eff}$ is not induced. The superconductivity is only suppressed by the leakage of the Cooper pairs into the F layer. At the same time for OOP orientation both suppressing factors work. It is further supported by the direct study of the the effective exchange field $h_{eff}$. Unlike our toy model, for the $\mathrm{NbSe_2/VSe_2}$ bilayer due to the anisotropy of the electronic spectra and presence of several Fermi surface branches the electronic spectra and DOS cannot be characterized by a single unified parameter $h_{eff}$. The value of the effective exchange field depends on the particular Fermi surface and on the momentum direction. To investigate this dependence we calculated the partial DOS for electrons in the vicinity of a given Fermi surface and for a given momentum direction. It is calculated according to Eq.~(\ref{DOS}), where the momentum integration is performed not over the whole BZ, but over the small circular momentum region with the radius $p_l \sim 0.3a^{-1}$ (with $a$ being the lattice parameter of the heterostructure) in the vicinity of the corresponding Fermi surface and for the given momentum direction. 

The effective exchange field $h_{eff}^{\Gamma (\mathrm K)}(\bm n)$ for a given momentum direction $\bm n = \bm p/|\bm p|$ and for a given Fermi surface around the $\Gamma (\mathrm K)$-point is extracted from the Zeeman splitting of the coherence peaks of this partial DOS. More precisely, we consider the sum of the partial DOS corresponding to $\bm n$ and $-\bm n$ because this symmetrization procedure eliminates the spin splitting  originated from the Ising-type SOC. It is obtained that for IP orientation  $h_{eff}^{\Gamma (K)}(\bm n) \approx 0$. For OOP orientation it is presented in Figs.~\ref{fig:real_zeeman}(b) and (c) for two different $V$. In Fig.~\ref{fig:real_zeeman}(c) corresponding to $V=550$meV it is clearly seen that in the regions, where $\mathrm{NbSe_2}$ and $\mathrm{VSe_2}$ Fermi surfaces intersect and, consequently, the hybridization of the electronic spectra is strong, $h_{eff}^{\Gamma (\mathrm K)}(\bm n)$ has sharp peaks of different signs. In other regions of the $\mathrm{NbSe_2}$ Fermi surfaces, which are not affected by the strong hybridization, the value of $h_{eff}^{\Gamma (\mathrm K)}(\bm n)$ is smaller. As a result, the behavior of $h_{eff}^{\Gamma (\mathrm K)}(\bm n)$ becomes strongly anisotropic. On the contrary, Fig.~\ref{fig:real_zeeman}(b) represents the case with no intersections of the $\mathrm{NbSe_2}$ and $\mathrm{VSe_2}$ Fermi surfaces and, consequently, weaker hybridization. We can see that in this case the sharp peaks in $h_{eff}^{\Gamma (\mathrm K)}(\bm n)$ are absent, the behavior of this quantity in the BZ is more isotropic. 

Fig.~\ref{fig:real_zeeman}(a) demonstrates $h_{eff}^{\Gamma(\mathrm K)}$ extracted from the splitting of the coherence peaks of the DOS averaged over all momentum directions at the Fermi surface around $\Gamma$-point (sum around all $\mathrm K$-points) as a function of $V$. The sign-changing behavior of $h_{eff}^{\Gamma}$ is similar to the behavior obtained in our toy model and presented in Fig.~\ref{fig:gating_mu}. It is not surprising because this Fermi-surface is nearly circular and is affected by the Ising-type SOC only weakly. The behavior of $h_{eff}^{\mathrm K}$ is not similar to the behavior obtained in the framework of the toy model because of the strong anisotropy and strong Ising-type SOC.

\section{Conclusions}
In this work we unveil the microscopic physical mechanism of the magnetic proximity effect, that is the suppression of the superconductivity by the exchange field of the adjacent ferromagnet in 2D S/F  vdW heterostructures and predict that  it is determined by the degree of hybridization of electronic spectra of the individual materials. The degree of hybridization can be adjusted by changing the  relative filling factors by gating one of the materials allowing for high controllability of the magnetic proximity effect. We illustrate the underlying physics of the processes governing the proximity effect using a minimal tight-binding hamiltonian model on a square lattice, and then demonstrate the existence of the same effects in heterostructures based on vdW materials: a monolayer 1H-$\mathrm{NbSe_2}$ as  a superconductor and a monolayer 1T-$\mathrm{VSe_2}$ as a ferromagnet. It is demonstrated that the Zeeman splitting of the DOS can be switched on/off and reversed by gating and nonmonotonic dependencies of the supercondicting order parameter on the gating potential are obtained.

\begin{acknowledgments}
G.A.B. and I.V.B. acknowledge the support from Theoretical Physics and Mathematics Advancement Foundation “BASIS” via the project No. 23-1-1-51-1 and from MIPT via Project No. FSMG-2023-0014. K.A.B. acknowledges Saint-Petersburg State University for a research project No. 95442847. The calculations performed for $\mathrm{NbSe_2/VSe_2}$ structures were supported by the Russian Science Foundation via the RSF project No. 24-12-00152.
M.M.O. acknowledges the support by MCIN/AEI/10.13039/501100011033/
(Grant No. PID2022-138210NB-I00) and “ERDF A way of making Europe,” by Ayuda CEX2023-001286-S financiada por MICIU/AEI/10.13039/501100011033, as well as MCIN with funding from European Union NextGenerationEU (PRTR-C17.I1) promoted by the Government of Aragon.
\end{acknowledgments}

\appendix

\section{Green's functions approach to 2D S/F bilayer}

\label{app:A}

Here we describe the Gor'kov Green's function method, which we formulate and use to calculate observables in the considered bilayer vdW heterostructure.  We begin with the following tight-binding Hamiltonian:
\begin{align}
&\hat H = \sum\limits_{i,\alpha,\beta}\hat c^\dagger_{i,\alpha}\left(\begin{matrix}0&0\\0&(\bm h\bm\sigma)_{\alpha,\beta}\\\end{matrix}\right)\hat c_{i,\beta}-
\sum\limits_{i,\sigma}\hat c^\dagger_{i,\sigma}\left(\begin{matrix}\mu_S&0\\0&\mu_F\\\end{matrix}\right)\hat c_{i,\sigma}\nonumber \\ 
&+\sum\limits_i\left[\hat c_{i,\uparrow}\left(\begin{matrix}\Delta&0\\0&0\\\end{matrix}\right)\hat c_{i,\downarrow}+H.c.\right] \nonumber \\
&- \sum\limits_{<ij>,\sigma}\hat c^\dagger_{i,\sigma}\left(\begin{matrix}t_S^{ij,\sigma}&0\\0&t_F^{ij,\sigma}\\\end{matrix}\right)\hat c_{j,\sigma}-\sum\limits_{i,\sigma}\hat c^\dagger_{i,\sigma}\left(\begin{matrix}0&t_{SF}\\t_{SF}&0\\\end{matrix}\right)\hat c_{i,\sigma}~~~~~~~
\label{eq:hamiltonian_gen}
\end{align}
It is a generalized version of Hamiltonian (\ref{eq:hamiltonian}), where to correctly describe real materials we assume not only nearest-neighbor hopping, but arbitrary hops $i\to j$ with complex hopping elements $t^{ij,\sigma}_S,t^{ij,\sigma}_F$ in the S and the F layers, respectively. In general, the hopping elements can depend on spin and their phases account for the spin-orbit interaction. $\langle ij \rangle$ means summation over all involved neighbors. 

The Matsubara Green's function in the two-layer formalism is $8 \times 8$ matrix in the direct product of spin, particle-hole and layer spaces. Introducing the two-layer Nambu spinor $\check \psi_{\bm i} = (\hat c_{{\bm i},\uparrow}^S, \hat c_{\bm i,\downarrow}^S, \hat c_{\bm i,\uparrow}^F,\hat c_{\bm i,\downarrow}^F, \hat c_{\bm i,\uparrow}^{S\dagger}, \hat c_{\bm i,\downarrow}^{S\dagger}, \hat c_{\bm i,\uparrow}^{F\dagger}, \hat c_{\bm i,\downarrow}^{F\dagger})^T$ we define the Green's function as follows: 

\begin{eqnarray}
\check G_{\bm i \bm j}(\tau_1, \tau_2) = -\langle T_\tau \check \psi_{\bm i}(\tau_1) \check \psi_{\bm j}^\dagger(\tau_2) \rangle,
\label{Green_Gorkov}
\end{eqnarray}
where $\langle T_\tau ... \rangle$ means  imaginary time-ordered thermal averaging. Introducing Pauli matrices in spin, particle-hole and layer spaces: $\sigma_k$, $\tau_k$ and $\rho_k$ ($k=0,x,y,z$) and operator $\hat j $ as
\begin{eqnarray}
\hat j \check c_{\bm i} = \sum\limits_{<ij>,\sigma}\left(\begin{matrix}t_S^{ij,\sigma}&0\\0&t_F^{ij,\sigma}\\\end{matrix}\right)\hat c_{j,\sigma}
\label{op_j}
\end{eqnarray}
one can obtain the Gor'kov equation for Green's function in terms of the Matsubara frequencies $\omega_m = \pi T(2m+1)$. The derivation is similar to that described in Ref.~\onlinecite{Bobkov2022}. The resulting Gor'kov equation takes the form:
\begin{align}
G_{\bm i}^{-1} \check G_{\bm i \bm j}(\omega_m) = \delta_{\bm i \bm j}, \label{gorkov_eq_ml}
\end{align}
\begin{align}
G_{\bm i}^{-1} = \tau_z \left( \hat j + \left(\begin{matrix}\mu_S&0\\0&\mu_F\\\end{matrix}\right) - \frac{\rho_0+\rho_z}{2}\check \Delta_{\bm i} i \sigma_y \right. \nonumber \\
\left. - \bm h \check {\bm \sigma} \delta_{i_x,0} \frac{\rho_0-\rho_z}{2} + \left(\begin{matrix}0&t_{SF}\\t_{SF}&0\\\end{matrix}\right)\right) + i \omega_m .
\label{G_i}
\end{align}
where $\check \Delta_{\bm i} = \Delta_{\bm i} \tau_+ + \Delta_{\bm i}^* \tau_-$ with $\tau_\pm  = (\tau_x \pm i \tau_y)/2$ and $\check {\bm \sigma} = \bm \sigma (1+\tau_z)/2 + \bm \sigma^* (1-\tau_z)/2$ is the quasiparticle spin operator. Further we consider the Green's function in the mixed representation:
\begin{eqnarray}
\check G(\bm R, \bm p) = F(\check G_{\bm i \bm j}) = \int d^2 r e^{-i \bm p(\bm i - \bm j)}\check G_{\bm i \bm j},
\label{mixed}
\end{eqnarray}
where $\bm R=(\bm i+\bm j)/2$ and the integration is over $\bm i - \bm j$. Now we define the following  transformed Green's function to simplify further calculations and to present the Gor'kov equation in a more common form:

\begin{eqnarray}
\check {\tilde G}(\bm R, \bm p) = 
\left(
\begin{array}{cc}
1 & 0 \\
0 & -i\sigma_y
\end{array}
\right)_\tau  \check G(\bm R, \bm p)  
\left(
\begin{array}{cc}
1 & 0 \\
0 & -i\sigma_y
\end{array}
\right)_\tau ,
\label{unitary}
\end{eqnarray}
where and below subscript $\tau$($\rho$) means that the explicit matrix structure corresponds to the particle-hole (layer) space. Then we obtain:

\begin{align}
    \left( 
    i\omega_m\tau_z-\left(\begin{matrix}\hat \xi_S(\bm p)&0\\0&\hat \xi_F(\bm p)\\\end{matrix}\right)_\rho \right. \nonumber \\
    \left. +\left(\begin{matrix}\tau_z \check \Delta &0\\0&-\bm h\bm \sigma \tau_z\\\end{matrix}\right)_\rho+t_{SF}\rho_x
    \right) \check {\tilde G}(\bm p)=1
\end{align}
where $\hat \xi_{S,F}$ are diagonal matrices in spin space, describing normal state electron spectra of the S and F layers, with elements $\xi_{S,F}^\sigma(\bm p)=-\sum\limits_{<0j>,\sigma} t_{S,F}^{0j,\sigma}e^{i\bm p \bm j }-\mu_{S,F}$. Here we omit the spatial coordinate $\bm R$ in the Green's function, the order parameter and the exchange field due to the translational invariance along the S/F interface.

The superconducting order parameter of the S/F bilayer is calculated from the self-consistency equation
\begin{eqnarray}
\Delta =  \lambda  T \sum \limits_{\omega_m}\int \frac{d^2p}{(2\pi)^2} \frac{{\rm Tr}[\check {\tilde G}\sigma_0(\tau_x-i\tau_y)(\rho_0+\rho_z)]}{8} , ~~
\label{SC}    
\end{eqnarray}
Electron spectral function can be calculated as
\begin{eqnarray}
A_{S,F} =   \frac{{\rm Tr}[\check {\tilde G}^R(\sigma_0+s \sigma_z)(\tau_0+\tau_z)(\rho_0\pm\rho_z)]}{8} ,
\label{A}    
\end{eqnarray}
where $A_{S,F}$ is a function of $(\bm p, \varepsilon, s)$ with $s=\pm 1$ and $\varepsilon$ being spin and energy of the electron, respectively. $\check {\tilde G}^R$ can be obtained from $\check {\tilde G}$ with substituting $i\omega_m \to \varepsilon+i\delta$, $\delta$ is a positive infinitesimal imaginary part. DOS is the momentum-integrated spectral function:
\begin{eqnarray}
N_{S,F}(\bm \varepsilon, s)=\int \frac{d^2 p}{(2\pi)^2} A_{S,F}(\bm p, \bm \varepsilon, s)
\label{DOS}    
\end{eqnarray}

\section{DFT calculations }

\label{app:B}

\begin{figure}[tb]
	\begin{center}
		\includegraphics[width=85mm]{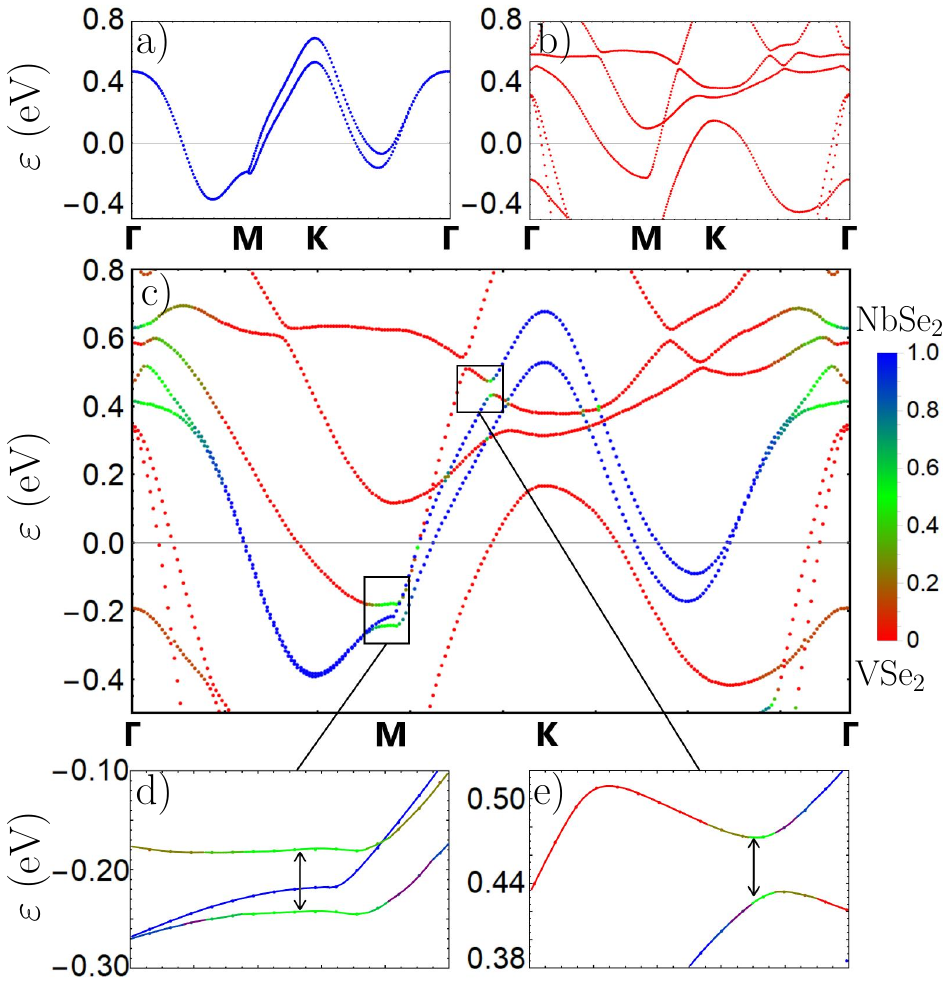}
\caption{(a)-(b) DFT-calculated low-energy band structures of the (a) 1H-$\mathrm{NbSe_2}$ monolayer in a nonsuperconducting state, (b) 1T-$\mathrm{VSe_2}$ monolayer in the ferromagnetic state, and (c) $\mathrm{NbSe_2/VSe_2}$ heterostructure. In (c), the colors encode the spectral weights:  parts  of the heterostructure spectrum, which originate from the $\mathrm{NbSe_2}$($\mathrm{VSe_2}$) layer and are  not significantly affected by the hybridization are shown in blue (red), while those significantly modified by the hybridization between $\mathrm{NbSe_2}$ and $\mathrm{VSe_2}$ are shown in green. (d)-(e) Magnified views of the anticrossings, which appeared as a result of the NbSe$_2$/VSe$_2$ hybridization. }
 \label{fig:DFTspectra}
	\end{center}
 \end{figure}
This section describes the electronic structure calculations of the individual 1H-NbSe$_2$ and 1T-VSe$_2$ layers as well as the ${\rm NbSe_2 / VSe_2}$ heterostructure using the density functional theory (DFT). All  calculations were performed using OpenMX package \cite{Ozaki2003},\cite{Ozaki2004} within the Perdew-Burke-Ernzerhof (PBE) generalized gradient approximation (GGA) \cite{Perdew1996}. 
Electronic spectra were calculated taking spin-orbit coupling into account.
An effective on-site Hubbard $U$ term of 1.0 eV was used to treat the V 3d electrons within the DFT+$U$ approach \cite{Anisimov1991, Ryee_2018} At the specified $U$ value VSe$_2$ is ferromagnetic, in agreement with the available experiments.
DFT-D3 method \cite{Grimme2010} was employed to take into account vdW interactions during the structure optimizations. 

To simulate both the individual monolayers and bilayer heterostructure the supercell approach was used. The thickness of the vacuum layer was larger than 20 ${\rm \AA}$  to minimize interactions between periodic images.
The energy cutoff of 3000 eV was used for the numerical integrations. 
2D Brillouin Zone was sampled by the $\Gamma$-centered $14\times14\times1$ k-point grid. 
To match 1H-NbSe$_2$ and 1T-VSe$_2$ in the $(1\times1)$ cell the lattice parameter of the heterostructure was set to $a=3.40{\rm \AA}$, which is the average of the experimental lattice parameters of the single layers, $a_{\rm NbSe_2}=3.45{\rm \AA}$ and $a_{\rm VSe_2}=3.35{\rm \AA}$ \cite{Yu2019}.
The positions of atoms were optimized until the ionic forces were smaller than 0.005 eV/${\rm \AA}$. The electronic convergence criterion was set to $3\cdot 10^{-6}$eV. The pseudo-atomic orbitals with cutoff radius $7{\rm \AA}$ were used in calculations.

For the calculations in the framework of the Green's functions approach, described in the previous section, we need the electron dispersions $\xi_{S,F}^\sigma$ of  separate $\mathrm{NbSe_2}$ and  $\mathrm{VSe_2}$ monolayers and the interlayer hopping element $t_{SF}$. To obtain $\xi_{S,F}^\sigma$ the DFT-calculated low-energy electron spectra $\varepsilon_{S,F}^\sigma$ were fitted by a single-band tight-binding model in a triangular lattice, taking into account the complex hopping elements between first to sixth
neighbours $t_0 e^{i\varphi_0} - t_5 e^{i\varphi_5}$, where $t_i$ is the corresponding hopping energy and $\varphi_i$ accounts for the spin-orbit coupling.  Only $\varphi_0 \neq 0$ for the case of the Ising-type spin-orbit coupling, which occurs in the considered materials.  The fitting formula takes the form\cite{Aikebaier2022}:
\begin{widetext}
\begin{align}\nonumber
    \varepsilon_{S,F}^\sigma = -2t_0[\cos(2\alpha-\varphi_0\sigma)+2\cos\beta\cos(\alpha+\varphi_0\sigma)]-2t_1[\cos(3\alpha-\beta)+~~~~~\\ \nonumber
+\cos(2\beta)+\cos(3\alpha+\beta)]-2t_2[\cos(2\alpha+2\beta)+\cos(2\alpha-2\beta)+\cos(4\alpha)]+~~~ \\ 
    -2t_3[\cos(5\alpha+\beta)+\cos(4\alpha-2\beta)+\cos(4\alpha+2\beta)+\cos(\alpha-3\beta)+\cos(5\alpha-\beta)+~\\ \nonumber
    \cos(\alpha+3\beta)]-2t_4[\cos(3\alpha-3\beta)+\cos(3\alpha+3\beta)+\cos(6\alpha)]-t_5[\cos(6\alpha-2\beta)+ \\ \nonumber
    +\cos(4\beta)+\cos(6\alpha+2\beta)]-\mu+h\sigma = \xi_{S,F}^\sigma + \sigma h~~~~~~~~~~~~~~~~~~~~
\end{align}
where $\alpha=p_x a/2$, $\beta=\sqrt{3}p_y a/2$, $\mu$ is the chemical potential, and $h$ is the exchange splitting, $\sigma=\pm 1$ means the spin of the electron. 

Fig. \ref{fig:DFTspectra} shows the DFT-calculated electron spectra of the single layers and that of the heterostructure. The parameters extracted from the fits of the data presented in Fig.~\ref{fig:DFTspectra} are listed in Table~\ref{tab:hopping_par}. For the ${\rm NbSe_2}$ they are in good agreement with the data reported earlier \cite{Aikebaier2022}. 
Anticrossings at the "intersections" of the ${\rm NbSe_2}$ and ${\rm VSe_2}$ spectra allow us to estimate the interlayer hopping element $t_{SF}$ because the gap due to an anticrossing is equal to $2t_{SF}$. We obtain that $t_{SF}$ can reach up to $30$ meV, which for the studied here case of the ideally aligned and matched ${\rm NbSe_2}$ and ${\rm VSe_2}$ lattices should be considered as an upper limit estimate. It is worth noting that the anticrossing region presented in Fig.~\ref{fig:DFTspectra}(d) corresponds to the band approximated by our tight-binding hamiltonian. At the same time, the anticrossing region presented in Fig.~\ref{fig:DFTspectra}(e) involves another branch of the ${\rm VSe_2}$ spectrum, but the estimated value of the interlayer hopping is the same for this anticrossing point.

\begin{table}
\begin{center}
\begin{tabular}{|c|c|c|c|c|c|c|c|c|c|}
\hline
 & $\mu$, meV & $t_0$, meV & $t_1$, meV & $t_2$, meV & $t_3$, meV & $t_4$, meV & $t_5$, meV & $\varphi_0$ & $h$, meV \\
\hline
${\rm NbSe_2}$ & 31.4 & 17.5 & 99.8 & -7.8 & -3.6 & -14.3 & 0.5 & 1.48 & 0 \\
\hline
${\rm VSe_2}$ & -18.8 & -22.2 & 93.4 & -65.4 & 17.3 & -23.6 & 8.1 & 0.2 & 401 \\
\hline
\end{tabular}
\end{center}
\caption{\label{tab:hopping_par}Parameters of the one-band tight-binding model fitted to the DFT-calculated electron spectra.}
\end{table} 

\end{widetext}

\bibliography{vdW}

\end{document}